\documentclass[twocolumn, 10pt]{tsfp7}
\usepackage{flushend}
\usepackage[pdftex]{graphicx}
\usepackage{amsmath}

\title{On the dissipation rate coefficient in homogeneous isotropic decaying and forced turbulence}

\author{P. Schaefer
    \affiliation{
	Department of Combustion Technology \\
	RWTH Aachen University\\
	Templergraben 64, 52056 Aachen, Germany\\
    pschaefer@itv.rwth-aachen.de
    }	
}

\begin{document}

\maketitle   

\fontsize{9}{11}\selectfont

\section*{ABSTRACT}
%
%
%
The normalized non-dimensional von K\'arm\'an-Howarth equation for isotropic homogeneous decaying and forced steady turbulence is integrated to obtain expressions for the dissipation rate coefficient $C_{\epsilon}=(L \epsilon)/\langle u^2 \rangle^{3/2}$, where $L$ denotes the longitudinal integral length scale, $\epsilon$ the mean dissipation rate and $\langle u^2 \rangle$ the mean variance of the longitudinal velocity fluctuations. For decaying turbulence the final exact expressions for $C_{\epsilon}$ for the low and high Reynolds number limit depend on the decay exponent $n$, which is known to depend on the initial velocity structure at the turbulence production. The dependence on $n$ leads to a non-universal coefficient. The expressions for the steady forced case depend on the forcing mechanism and thus are not universal either. Nonetheless, a lower value and considerably less scatter as compared to the decaying turbulence case should be expected when similiar forcing algorithms are employed. 
%
%
%

\section*{INTRODUCTION}
%

A fundamental assumption on which the phenomenology of turbulence is based, is the independence of the mean energy dissipation $\epsilon$ of the viscosity $\nu$, provided the latter is very small. This concept, also termed dissipation anomaly, cf. Frisch \cite{Frisch95}, has first been introduced by Taylor \cite{Taylor35} in 1935 and it directly translates, based on dimensional grounds, to a proportionality between $\epsilon$ and $\langle u^2 \rangle^{3/2}/L$ with $L$ being an integral length scale. The assumption of a universal proportionality constant plays a fundamental role in turbulence modeling and even directly determines one of the model constants in today's widely used one- and two-equation RANS models, such as the $k - \varepsilon$ model or the $k - \omega$ model, where the unknown Reynolds stresses are closed based on an eddy viscosity ansatz. 
However, as early as 1953 when Batchelor \cite{Batchelor53} plotted values of the proportionality constant in his textbook the question arose, whether or not it is truly universal and if it is not, how it depends on the flow configuration under consideration. 
Experimental data from various types of decaying turbulent flows such as wakes, jets and grid turbulence (with different geometries) have been analyzed and the resulting values of $C_{\epsilon}$ have been collected for instance by Sreenivasan \cite{Sreenivasan84,Sreenivasan97} and Burattini \cite{Burattini05}. The large scatter of the data, even for similar flow configurations, has led all of the above mentioned authors to the conclusion that the constant is not universal and depends critically on the way the turbulence is generated. Recently Goto and Vassilicos \cite{Goto09} showed that the value of $C_{\epsilon}$ depends on the internal stagnation point structure and can thus not be universal, as the stagnation points themselves depend on the large scales of the flow. The purpose of this work is to derive expressions for $C_{\epsilon}$ based on the von K\'arm\'an-Howarth equation, for the case of homogeneous isotropic decaying turbulence and homogeneous isotropic forced steady turbulence. The paper is organized as follows. In Chapter 1 we will derive exact expressions for the dissipation rate coefficient for decaying and forced turbulence based on the normalized von K\'arm\'an-Howarth equation. The results of the analysis are discussed and compared with each other in chapter 2. In chapter 3 a distinction is made between the limit of large and small Reynolds numbers and exact and approximate results for the dissipation rate coefficient are derived for both cases. Concluding remarks are given in chapter 4. 
\section{ANALYSIS}

For isotropic homogeneous turbulence we define the non-dimensionalized two-point longitudinal second and third order structure functions as 

\begin{equation}
\begin{split}
& d_{LL}(r,t)=  \frac{\langle \left( u(x+r,t) - u(x,t) \right)^2 \rangle}{\langle u^2 \rangle},\\
& d_{LLL}(r,t)=  \frac{\langle \left( u(x+r,t) - u(x,t) \right)^3 \rangle}{\langle u^2 \rangle^{\frac{3}{2}}}.
\label{dlldlll2}
\end{split}
\end{equation}

where $u(x,t)$ denotes the instantaneous longitudinal velocity fluctuation at point $x$ which is seperated by a distance $r$ from the second point. The non-dimensional second order structure function can be transformed into the non-dimensional two-point correlation yielding

\begin{equation}
f(r,t)=\frac{\langle u(x+r)u(x)\rangle}{\langle u^2 \rangle} = 1 - \frac{d_{LL}}{2},
\label{f}
\end{equation}

where angular brackets denote ensemble averages.

Let us define two length scales 

\begin{equation}
\begin{split}
& L(t)=\int_0^{\infty}f(r,t)dr,\\
& l_t(t)=\frac{\langle u^2 \rangle^{3/2}}{\epsilon},
\end{split}
\label{Lintegral}
\end{equation}

%

the ratio of which is precisely the above introduced energy dissipation rate coefficient as 

\begin{equation}
C_{\epsilon}=\frac{L}{l_t}=\frac{L \epsilon}{\langle u^2 \rangle^{3/2}}.
\label{Cepsilon}
\end{equation}

\subsection{Decaying turbulence}

We write the von K\'arm\'an-Howarth equation, cf. von K\'arm\'an and Howarth \cite{Karman38}, following an approach by Lundgren \cite{Lundgren02} as 

\begin{equation}
\begin{split}
& -\frac{2}{3} \epsilon f(r,t) + u^2 \frac{\partial f(r,t)}{\partial t}= \\& \frac{2}{3} \langle u^2 \rangle^{3/2} \left( \frac{d_{LLL}(r,t)}{r} + \frac{1}{4} \frac{\partial d_{LLL}(r,t)}{\partial r} \right)+ \\
& 2 \nu u^2 \left( \frac{\partial^2 f(r,t)}{\partial r^2} + \frac{4}{r} \frac{\partial f(r,t)}{\partial r} \right),\\
\end{split}
\label{VKH}
\end{equation}

where $\nu$ denotes the molecular viscosity of the fluid.

Let us define the new coordinates

\begin{equation}
\tilde{r} = r/L(t), \;\;\; \tau = t.
\label{coord}
\end{equation}
Then the time derivative $\partial f(\tilde{r},t) / \partial t$ in eq. \ref{VKH} becomes

\begin{equation}
\frac{\partial f(\tilde{r},t)}{\partial t} = \frac{\partial f(\tilde{r}, \tau)}{\partial \tau} + \frac{\partial \tilde{r}}{\partial t} \frac{\partial f(\partial \tilde{r}, \tau)}{\tilde{r}}.
\label{dfdt}
\end{equation}

The derivative $\partial \tilde{r}/ \partial{t}$ may be expressed as

\begin{equation}
\frac{\partial \tilde{r}}{\partial t} = - \frac{\tilde{r}}{L} \frac{d L}{dt}.
\label{dvdt}
\end{equation}

Introducing the new coordinates into eq. \ref{VKH} (and dropping the functions' arguments for simplicity) yields 

\begin{equation}
\begin{split}
-\underbrace{\frac{\epsilon L}{\langle u^2 \rangle^{3/2}}}_{C_\epsilon^{d}} f + \frac{3}{2} L \langle u^2 \rangle^{-1/2} \left( \frac{\partial f}{\partial \tau} - \frac{1}{L} \frac{dL}{dt} \tilde{r} \frac{\partial f}{\partial \tilde{r}}  \right) \\= \left( \frac{d_{LLL}}{\tilde{r}} + \frac{1}{4} \frac{\partial d_{LLL}}{\partial \tilde{r}} \right) +  \frac{3}{Re_L} \left( \frac{\partial^2 f}{\partial \tilde{r}^2} + \frac{4}{\tilde{r}} \frac{\partial f}{\partial \tilde{r}} \right), 
\label{VKHfinal}
\end{split}
\end{equation}

(the superscript d stands for decaying) with $Re_L=(\langle u^2 \rangle^{1/2} L)/\nu$ a Reynolds number based on the integral length scale defined in eq. \ref{Lintegral}.

Noting that by definition $\int_0^{\infty} f d \tilde{r} = 1$ we integrate eq. \ref{VKHfinal} to obtain

\begin{equation}
\begin{split}
& -C_{\epsilon}^{d} + \frac{3}{2} L \langle u^2 \rangle^{-1/2} \underbrace{\frac{\partial }{\partial \tau} \int_0^{\infty} f d \tilde{r}}_{=0} -\frac{3}{2} \langle u^2 \rangle^{-1/2} \frac{dL}{dt} \underbrace{\int_0^{\infty} \tilde{r} \frac{\partial f}{\partial \tilde{r}} d \tilde{r}}_{=-1}  = \\& \int_0^\infty \frac{d_{LLL}}{\tilde{r}} d\tilde{r} + \frac{1}{4} \underbrace{\int_0^\infty \frac{\partial d_{LLL}}{\partial \tilde{r}} d\tilde{r}}_{=0} \\& +
\frac{3}{Re_L}  \left(\underbrace{\int_0^\infty \frac{\partial^2 f}{\partial \tilde{r}^2} d\tilde{r}}_{=0}+ \int_0^\infty \frac{4}{\tilde{r}} \frac{\partial f}{\partial \tilde{r}} d \tilde{r} \right),
\label{intVKH}
\end{split}
\end{equation}

where the second integral on the left hand side is obtained by partial integration. Eq. \ref{intVKH} thus simplifies to

\begin{equation}
-C_{\epsilon}^{d} + \frac{3}{2} \langle u^2 \rangle^{-1/2} \frac{dL}{dt} = \int_0^{\infty} \frac{d_{LLL}}{\tilde{r}} d \tilde{r} + \frac{12}{Re_L} \int_0^\infty \tilde{r}^{-1} \frac{\partial f}{\partial \tilde{r}} d\tilde{r}.
\end{equation}

Using relation \ref{Cepsilon} we write $L=C_{\epsilon} \langle u^2 \rangle^{3/2}/ \epsilon$ so that 

\begin{equation}
\frac{dL}{dt} = C_{\epsilon}^{d} \frac{d}{dt} \left( \frac{\langle u^2 \rangle^{3/2}}{\epsilon} \right) = C_{\epsilon}^{d} \left( \frac{3/2 \langle u^2 \rangle^{1/2} \frac{d u^2}{dt} \epsilon - \langle u^2 \rangle^{3/2} \frac{d \epsilon}{dt}}{\epsilon^2} \right).
\end{equation}

with $d u^2/dt = - 2/3 \epsilon$ for isotropic decaying turbulence. Under the assumption that the turbulence has reached an asymptotic state at which the decay of the turbulent kinetic energy follows a power law decay of the form $k(t) \propto t^{-n}$ with a constant decay exponent $n$, which corresponds to self-preserving solutions of the von K\'arm\'an-Howarth equation (for a detailed discussion see Dryden \cite{Dryden43}, Batchelor \cite{Batchelor48}, Korneyev \cite{Korneyev76}), one obtains 

\begin{equation}
-C_{\epsilon}^{d} \left(\frac{5}{2} + \tau_i \frac{d \ln \epsilon}{dt} \right) = \int_0^{\infty} \frac{d_{LLL}}{\tilde{r}} d\tilde{r} + \frac{12}{Re_L} \int_0^\infty \tilde{r}^{-1} \frac{\partial f}{\partial \tilde{r}} d\tilde{r},
\label{almostfinal}
\end{equation}

where $\tau_i=k/\epsilon$ defines an integral eddy turn over time scale. Eq. \ref{almostfinal} can be further simplified as $\tau_i = t/n$ and $d \ln \epsilon/dt = -(n+1)/t
$, so that

\begin{equation}
\tau_i \frac{d \ln \epsilon}{dt} =- \left( 1 + \frac{1}{n}\right),
\end{equation}

yielding

\begin{equation}
C_{\epsilon}^{d}   = \left( \frac{2n}{3n-2} \right) \left( \int_0^{\infty} \frac{(-d_{LLL})}{\tilde{r}} d\tilde{r} + \frac{12}{Re_L} \int_0^\infty \tilde{r}^{-1} \frac{\partial f}{\partial \tilde{r}} d\tilde{r}\right).
\label{final1}
\end{equation}

Eq. \ref{final1} constitutes an exact relation for homogeneous isotropic decaying turbulence. At this point it is pertinent to make reference to Rotta's \cite{Rotta72} work, who used Heisenberg's \cite{Heisenberg48} theory in combination with Loitsianskii's \cite{Loitsyanski39} invariant hypothesis to derive explicit expressions for $C_\epsilon^{d} $ for the low and high Reynolds number limit of decaying turbulence. Different from Rotta's approach, the above analysis does not assume any invariant integral of motion or ad-hoc assumptions and the final exact expression (eq. \ref{final1}) preserves the explicit dependence on the decay exponent. 

\subsection{Forced turbulence}
 
In the case of forced turbulence we write the von K\'arm\'an-Howarth equation following an approach by Fukayama \cite{Fukayama00} as

\begin{equation}
\begin{split}
\frac{\partial D_{LL}}{\partial t}+\frac{1}{3}r^{-4}\frac{\partial}{\partial r}\left(r^{4} D_{LLL}\right)=2\nu r^{-4}\frac{\partial}{\partial r}\left(r^4\frac{\partial D_{LL}}{\partial r}\right)\\+\frac{4}{3}(P-\epsilon)-4P F(r;l_f),
\label{VKHforced}
\end{split}
\end{equation} 
 
where $P$ denotes the mean energy input through the forcing and the last term on the right hand side of eq. \ref{VKHforced} is the forcing-velocity correlation term with

\begin{equation}
F(r;l_f)=r^{-3}\int^{r}_{0}s^{2} \cos\left(\frac{s}{l_f}\right)ds.
\end{equation}

Here $s$ is a dummy variable and the forcing is assumed to be random Gaussian, delta correlated in time, homogeneous and isotropic. Different from Fukayama \cite{Fukayama00} we assume the forcing to have compact support limited to the largest scale of the system denoted by $l_f$. Eq. \ref{VKHforced} is thus valid for $0 \leq r \leq l_f$. For a detailed derivation see Fukayama \cite{Fukayama00}. In the steady case we have

\begin{equation}
\frac{\partial D_{LL}}{\partial t}=0, \;\;\; P=\varepsilon,
\end{equation}

%

and the mean energy input must equal the mean dissipation. We write eq. \ref{VKHforced} in non-dimensional form using the coordinates defined in relation \ref{coord}, so that we obtain

\begin{equation}
\begin{split}
\frac{\partial d_{LLL}}{\partial\tilde{r}}+4\frac{d_{LLL}}{\tilde{r}}+\frac{12}{Re_{L}}\left(\frac{\partial^2 f}{\partial^2\tilde{r}}+4\tilde{r}^{-1}\frac{\partial f}{\partial \tilde{r}}\right)=-12C_{\epsilon}^{f} F(\tilde{r},l_f).
\end{split}
\end{equation}

Integration over the entire domain (where we implicitely let $l_f \rightarrow \infty$) yields 
\begin{equation}
\begin{split}
\int^{\infty}_{0}\frac{(-d_{LLL})}{\tilde{r}}d\tilde{r}+\frac{12}{Re_{L}}\int^{\infty}_{0}\tilde{r}^{-1}\frac{\partial f}{\partial \tilde{r}}d\tilde{r}=3C_{\epsilon}^{f} \int^{\infty}_{0}F\left(\tilde{r};l_f\right)d\tilde{r}.
\label{fastfinal_forced}
\end{split}
\end{equation}

The integral on the right hand side of eq. \ref{fastfinal_forced} can be solved analytically

\begin{equation}
\int^{\infty}_{0} F\left(\tilde{r};l_f\right)d\tilde{r} = \frac{l_f}{L} \left[ \sin(1) - \cos(1)\right] \approx 0.3 \frac{l_f}{L}.
\end{equation}

We thus obtain the equivalent expression of eq. \ref{final1} for the forced case,

\begin{equation}
\begin{split}
C_{\epsilon}^{f} \approx 1.1 \frac{L}{l_f} \left(\int^{\infty}_{0}\frac{(-d_{LLL})}{\tilde{r}}d\tilde{r}+\frac{12}{Re_{L}}\int^{\infty}_{0}\tilde{r}^{-1}\frac{\partial f}{\partial \tilde{r}}d\tilde{r}\right). 
\label{final_forced}
\end{split}
\end{equation}

\section{DISCUSSION}

Eq. \ref{final1} and eq. \ref{final_forced} constitute two exact results for the dissipation rate coefficient in homogeneous isotropic decaying and forced steady turbulence, respectively. A direct comparison of the two expressions reveals that they are identical in form and depend on the same integral expressions involving the non-dimensional third order structure function $d_{LLL}(\tilde{r})$ and the non-dimensional two-point correlation $f(\tilde{r})$. These integral expressions are only multiplied by a different factor, which in the case of decaying turbulence depends on the decay exponent $n$ and in the case of forced turbulence on the ratio of the longitudinal length scale $L$, defined in eq. \ref{Lintegral} to the scale $l_f$ at which energy is pumped into the system to sustain a steady turbulent flow. For the following discussion we assume the high Reynolds number limit of eq. \ref{final1} and eq. \ref{final_forced}, for which the last term on the right hand side proportional to $Re_L^{-1}$ can be neglected. In this limit the dissipation rate coefficient will be denoted as $C_{\varepsilon, \infty}$ and its expression only involves an integral over the non-dimensional third order structure function whose functional form is a-priori not known. As the upper integration limit extends to infinity one can deduce that the integrals involve contribution from the largest scales of the turbulent field which cannot be expected to be universal. For the case of decaying turbulence the decay exponent dependent factor can be expected to be different for different types of flows, as the decay exponent $n$ which is realized for the flow depends on the initial conditions as stated by George \cite{George89, George92} and Barenblatt \cite{Barenblatt96}. Experimental results from Ling \cite{LingWan72} have supported the non-universality of the decay exponent. This leads to the conclusion that the energy dissipation coefficient in the high Reynolds number limit cannot be universal for homogeneous isotropic decaying turbulence.
In the case of forced turbulence the value of the factor $L/l_f$ cannot be specified from first principles. Nonetheless, one can draw the valuable conclusion that it must be smaller than the decay exponent dependent term for decaying turbulence. Assuming $l_f$ to be of the size of the system (the computational domain for DNS calculations) then, as the non-dimensional two-point correlation $f(\tilde{r})$ decays to zero for $r \rightarrow l_f$ and $f(0)=1$, eq. \ref{Lintegral} leads to the conclusion that $L<l_f$, so that $L/l_f<1$, while for all possible values of the decay exponent, the coefficient $2n/(3n-2) \geq 1$. One must thus conclude that in the high Reynolds number limit the energy dissipation coefficient assumes smaller values for forced turbulence as compared to decaying turbulence. In addition the lack of a strongly varying parameter such as the decay exponent $n$ will lead to less scatter in the case of forced turbulence. 

\section{EXACT AND APPROXIMATE SOLUTIONS}

\subsection{Low Reynolds number limit}

For the low Reynolds number limit the dissipation rate coefficient is denoted by $C_{\epsilon, 0}^{d}$. This regime corresponds to the viscous dominated final period of decay which is also characterized by a power law decay and during which the contribution of the third order structure function in eq. \ref{VKH} can be neglected. The decay exponent however assumes values different from the ones of the high Reynolds number regime. As the von K\'arm\'an-Howarth equation is for this case in closed form, its general solution can be expressed in terms of a special function, the confluent hypergeometric function to be denoted by $M(\alpha,\beta, z)$, where $\alpha$ and $\beta$ are parameters and $z$ is the coordinate. The two-point correlation can then be written as

\begin{equation}
u^2 f=u^2 M(n,5/2,-\xi^2/8),
\label{confluent}
\end{equation}

where $n$ is the decay exponent and $\xi=r/\sqrt{\nu(t-t_0)}$ a similarity variable. $t_0$ is a virtual time which can be neglect against $t$ as the low Reynolds number limit in decaying turbulence corresponds to large times $t$. For a detailed discussion of the final period of decay see Barenblatt \cite{Barenblatt96}. 

Differentiating eq. \ref{confluent} with respect to $t$ yields

\begin{equation}
u^2 \frac{\partial f}{\partial t} - \frac{2}{3} \epsilon f = u^2 \frac{\partial M}{\partial \xi} \frac{\partial \xi}{\partial t} - \frac{2}{3} \epsilon M.
\label{confluent1}
\end{equation}

Dividing eq. \ref{confluent1} by $\langle u^2 \rangle^{3/2}$ and integrating yields,

\begin{equation}
\frac{1}{\langle u^2 \rangle^{1/2}} \frac{d L}{d t} - \frac{2}{3} C_{\epsilon,0}^{d}  = \frac{1}{\langle u^2 \rangle^{3/2}} \int_0^{\infty} \left( u^2 \frac{\partial M}{\partial \xi} \frac{\partial \xi}{\partial t} - \frac{2}{3} \epsilon M \right) dr.
\label{confluentint}
\end{equation} 

The left hand side of eq. \ref{confluentint} can be simplified to 

\begin{equation}
\frac{1}{\langle u^2 \rangle^{1/2}} \frac{d L}{d t} - \frac{2}{3} C_{\epsilon,0}^{d}  = C_{\epsilon,0}^{d} \left( \frac{2}{3n} - 1 \right).
\label{lhsconfluentint}
\end{equation} 

The first integral on the right hand side of eq. \ref{confluentint} can be simplified with 

\begin{equation}
dr = \sqrt{\nu t} d \xi, \; \; \; \frac{d \xi}{dt} = - \frac{\xi}{2t}
\end{equation}

%

to

\begin{equation}
\begin{split}
& \frac{1}{\langle u^2 \rangle^{1/2}} \int_0^{\infty} \frac{\partial M}{\partial \xi} \frac{\partial \xi}{\partial t} dr = - \frac{\nu^{1/2}}{2 \langle u^2 \rangle^{1/2} \sqrt{t}} \int_0^{\infty} \xi \frac{\partial M}{\partial \xi} d \xi =\\ & \frac{1}{2} \sqrt{\frac{2}{3n}} (C_{\epsilon,0}^{d})^{\frac{1}{2}} Re_L^{-1/2} \int_0^{\infty} M d \xi, 
\label{confluentint1}
\end{split}
\end{equation} 

where integration by parts and the fact that $t=3/2 n\; \langle u \rangle ^2/\varepsilon$ and $Re_{l_t}^{-1/2} = C_{\epsilon}^{1/2} Re_{L}^{-1/2}$ have been used. The second integral on the right hand side is more straightforward to manipulate and yields

\begin{equation}
-\frac{2}{3} \frac{\epsilon}{\langle u^2 \rangle^{\frac{3}{2}}} \int_0^{\infty} M dr = - \frac{2}{3} \sqrt{\frac{3n}{2}} (C_{\epsilon,0}^{d})^{\frac{1}{2}} Re_L^{-\frac{1}{2}} \int_0^{\infty} M d \xi.
\label{confluentint2}
\end{equation}

The final expression for the low Reynolds number limit reads

\begin{equation}
C_{\epsilon,0}^{d} = \left[ \frac{3n}{2-3n} \left(\sqrt{\frac{1}{6n}} -  \sqrt{\frac{2n}{3}} \right) \int_0^{\infty} M d \xi \right]^{2} Re_L^{-1}.
\label{cepsfinallow}
\end{equation}

Eq. \ref{cepsfinallow} is an exact expression for the final period of decay in homogeneous isotropic decaying turbulence. It obeys the well-known scaling $C_{\epsilon,0} \propto Re_L^{-1}$ while the proportionality constant depends on the decay exponent $n$ during the final period of decay (the integral over the confluent hypergeometric function is also fully determined by the value of $n$). 

Eq. \ref{cepsfinallow} constitutes an exact relation for which without further assumptions numerical results can be obtained and it reveals an explicite dependence on the decay exponent $n$ that governs this regime. The value of $n$ however depends, as it does in the high Reynolds number limit on the initial conditions of the turbulence production, cf. Barenblatt \cite{Barenblatt96}. As a consequence, the behaviour in the low Reynolds number limit cannot be expected to be universal. As stated by Barenblatt \cite{Barenblatt96}, solutions to the problem with a value of $n > 5/2$ are not of physical interest as they are structurally unstable with respect to the initial conditions while the range of possible values for $1 < n < 5/2 $ is continuous. The coefficient of the right hand side of eq. \ref{cepsfinallow} in front of $Re_{L}^{-1}$ can be calculated for a given value of $n$. Table \ref{coefficientsLow} shows this value for different values of $n$. Note that for $n=5/2$, which corresponds to the so called Loitsianskii case, the confluent hypergeometric function reduces to a Gaussian error function so that the integral over it can be expressed analytically as

\begin{equation}
\begin{split}
\int_0^{\infty} M(\frac{5}{2},\frac{5}{2},-\frac{\xi^2}{8})d \xi = \sqrt{2 \pi}.
\label{intgauss}
\end{split}
\end{equation}

The values obtained here are of the same order of magnitude as the ones reported by Bos \cite{Bos07} who derived an expression for the dissipation rate coefficient based on modelled energy spectra. For the final period of decay he finds for $C_{\varepsilon,0}^{d}\approx 19 .. 32 Re_{L}^{-1}$ depending on the assumptions made on the energy spectrum. 

\begin{table}
\centering
\begin{tabular}{lcccc}
\hline
\hline
n & 1 & 1.5 & 2.0 & 2.5\\
\hline
coefficient             &  $\approx 52.3$           & $\approx 20.4$       & $\approx 14.7$  & $\frac{480}{121} \pi \approx 12.5$     \\
\hline
\hline
\end{tabular}
\caption{Proportionality coefficient $C_{\epsilon,0}^{d} \propto Re_L^{-1}$ for the final period of decay.} 
\label{coefficientsLow}
\end{table}

\subsection{High Reynolds number limit}

Different from the low Reynolds number limit, no numerical values for the energy dissipation rate coefficient can be deduced from eqs. \ref{final1} or \ref{final_forced} without further knowledge of the form of $d_{LLL}(\tilde{r})$. One has thus to resort to a closure approximation for $d_{LLL}(\tilde{r})$ to solve the corresponding integrals. We will employ an eddy viscosity closure formulated in real space as proposed by Oberlack and Peters \cite{Oberlack93} of the form

\begin{equation}
d_{LLL}=-\alpha \tilde{r}\sqrt{d_{LL}}\frac{\partial d_{LL}}{\partial \tilde{r}} = -\frac{2}{3} \alpha \tilde{r} \frac{\partial d_{LL}^{3/2}}{\partial \tilde{r}}.
\label{eddyviscosity}
\end{equation}

In eq. \ref{eddyviscosity} the constant $\alpha=6/(5 C^{3/2})$ has been chosen to correctly reproduce Kolmorogov's four-fifth-law for the third order structure function $D_{LLL}=4/5 \left(\varepsilon r\right)$ and the two-third scaling for the second order structure function $D_{LL}=C \left( \varepsilon r \right)^{2/3}$ for small values of r, where $C$ is Kolmogorov's constant. Caution has to be exercised as such a closure assumption neglects the non-universal influence of the large scales on the functional form of $d_{LLL}(\tilde{r})$ and thus only one single value for the integral can be predict independent of the flow type. 

The integration of the right hand side of eq. \ref{final1} and eq. \ref{final_forced} yields

\begin{equation}
\begin{split}
\int_0^{\infty} \frac{(-d_{LLL})}{\tilde{r}}d \tilde{r} = \frac{2}{3} \alpha \int_0^{\infty} \frac{\partial d_{LL}^{3/2}}{\partial \tilde{r}} d\tilde{r} \\= \frac{2}{3}\alpha d_{LL}^{3/2}|_0^{\infty} = \frac{2^{5/2}}{3}\alpha=\frac{2^{7/2}}{5 C^{3/2}}.
\end{split}
\end{equation}

The expression for the dissipation rate coefficient in the high Reynolds number limit for decaying turbulence then reads

\begin{equation}
C_{\epsilon,\infty}^{d} = \frac{2n}{3n-2} \left(\frac{2^{7/2}}{5 C^{3/2}}\right).
\label{final}
\end{equation}

Sreenivasan \cite{Sreenivasan95} gives a compilation of values of Kolomogorov's constant $C$ taken from different experiments and finds $C=2.0 \pm 0.4$. Choosing these limits and varying the decay exponent $n$ in a range from 1.0 to 1.9 corresponding to the values observed in experiments and DNS calculations (cf. Ling and Wan \cite{LingWan72}, Uberoi \cite{Uberoi67} and Corrsin \cite{Corrsin74} and references therein), one obtains values of $C_{\epsilon, \infty}^{d}$ ranging from 0.6 to 2.2, which fall in the same range as the values quoted in the literature (cf. Sreenivasan \cite{Sreenivasan84,Sreenivasan97} and Burattini \cite{Burattini05}) and represent a large part of the scatter observed.

The expression for the dissipation rate coefficient in the high Reynolds number limit for forced steady turbulence reads

\begin{equation}
C_{\epsilon, \infty}^{f} \approx 1.1 \frac{L}{l_f} \left(\frac{2^{7/2}}{5 C^{3/2}}\right).
\label{forced_high}
\end{equation}

Different from the theory in the DNS calculations energy is pumped in over a finite band of large scales and not at a single scale $l_f$. To account for this difference, an equivalent, single wavenumber $\kappa_f$ is calculated, that corresponds to the band of forced scales $0 \leq \kappa \leq \kappa_0$ as

\begin{equation}
\kappa_f=\frac{\int_0^{\kappa_0} \kappa E(\kappa) d\kappa}{\int_0^{\kappa_0} E(\kappa) d \kappa}=\frac{q+1}{q+2} \kappa_0,
\label{forcingband}
\end{equation}

where the energy spectrum at the large scales is assumed to follow $E(\kappa) \propto \kappa^{q}$, cf. Goto \cite{Goto09}. For a calculation domain of size $2 \pi$ this yields

\begin{equation}
\frac{L}{l_f}=\frac{L \kappa_f}{2 \pi}=\frac{q+1}{q+2} \frac{L}{2 \pi} \kappa_0.
\label{forcingGoto}
\end{equation} 

Note that the approach taken in eq. \ref{forcingband} is only one possible way of determining a single forced scale and corresponds to calculating the center of gravity of the energy spectrum in the range $l_f < r < \infty$.

The ratio $L/l_f$ can for example be determined from the calculations carried out by Goto \cite{Goto09} and it turns out that it varies slightly between 0.42 and 0.49, confirming the hypothesis that $L<l_f$. With these values and varying the Kolmogorov constant in the above range the interval $0.28 \leq C_{\varepsilon, \infty}^{f} \leq 0.60$ can be deduced. 

Let us briefly discuss the above found values for $C_{\varepsilon, \infty}$ in the context of RANS turbulence modeling. One common approach to close the unknown Reynolds stresses in RANS turbulence models is to employ an eddy viscosity closure \cite{Pope00}, where assuming the energy dissipation coefficient to be universal in the high Reynolds number limit, one can show that the model constant $c_{\mu}$ is related to the energy dissipation coefficient, yielding $C_{\varepsilon, \infty} = \left(3/2 \right)^{3/2} c_{\mu}^{3/4}.$

%
%
%


With the commmon value of $c_{\mu}=0.09$ this leads to a value of $C_{\varepsilon, \infty}=0.30$ which is close to the lower bound predicted for forced turbulence and considerably lower than any value predicted for decaying turbulence. In the light of the non-universality of the energy dissipation coefficient the relation of the model constant to the theoretical energy dissipation coefficient is however rather hypothetical and the model constant cannot be predicted from first principles but rather an optimal value for different flow configuration has to be chosen.

\section{CONCLUSION}
An analysis based on the integration of the von K\'arm\'an-Howarth equation in the case of isotropic homogeneous decaying and steady forced turbulence has led to three exact relations for the energy dissipation rate coefficient $C_{\varepsilon}$. Of special interest is the high Reynolds number limit of these expressions and the question whether the energy dissipation rate coefficient attains a universal value for infinite Reynolds numbers. It could be concluded, that in the case of decaying turbulence no universal constant is attained as the final expression retains an explicite dependence on the decay exponent of the flow which is known to be non-universal. As for the steady forced case it is found that the final expression for $C_{\varepsilon}$ depends on the forcing mechanism. Although different forcing mechanisms are employed by different authors, it is concluded that, the values obtained from such DNS calculations will be close to each other when similar forcing schemes are employed and that they exhibit less scatter than in decaying turbulence, as a strongly varying parameter is absent. Based on the exact relations it could be concluded that the energy dissipation rate coefficient is smaller for forced turbulence that for decaying turbulence. Numerical values however cannot be obtained from first principles only. The introduction of a closure assumption for the third-order structure function allows the calculation of numerical values for both cases. These turn out to depend on the value of the Kolmogorov constant. For both decaying and forced turbulence a corridor of numerical values can be given which agrees fairly well with the values reported in the literature. 

\section*{ACKNOWLEDGEMENTS}
This work was founded by the Deutsche For\-schungs\-gemein\-schaft under Grant Pe 241/38-1. The author would like to thank Prof. Peters for his continuous support and all the opportunities offered. In addition the author would like to thank Dr. Mellado and Mr. Gampert for all the fruitful discussions.

\end{document}